\documentclass[12pt]{article}

\usepackage[UKenglish]{babel}

\usepackage{latexsym}
\usepackage{amssymb}
\usepackage{epsfig}

\begin{document}

\title{Anomalous localisation near the band centre in
the 1D Anderson model: Hamiltonian map approach}

\author{L. Tessieri${}^{1,2}$~\footnote{Corresponding author:
luca.tessieri@gmail.com}, I. F. Herrera-Gonz\'{a}lez${}^{3}$,
F. M. Izrailev${}^{4}$,\\
{\it ${}^{1}$ Istituto dei Sistemi Complessi} \\
{\it via Madonna del Piano, 10; I-50019 Sesto Fiorentino, Italy} \\
{\it ${}^{2}$ Instituto de F\'{\i}sica y Matem\'{a}ticas} \\
{\it Universidad Michoacana de San Nicol\'{a}s de Hidalgo} \\
{\it 58060, Morelia, Mexico} \\
{\it ${}^{3}$ Facultad de Ciencias F\'{\i}sico-Matem\'{a}ticas} \\
{\it Universidad Michoacana de San Nicol\'{a}s de Hidalgo} \\
{\it 58060, Morelia, Mexico} \\
{\it ${}^{4}$ Instituto de F\'{\i}sica, Universidad Aut\'{o}noma
de Puebla,} \\
{\it Puebla, 72570, Mexico}}

\date{12th December 2011}

\maketitle

\begin{abstract}
We present a full analytical solution for the localisation length in the
one-dimensional Anderson model with weak diagonal disorder in the vicinity
of the band centre. The results are obtained with the Hamiltonian map
approach that turns out to be more effective than other known methods.
The analytical expressions are supported by numerical data. We also discuss
the implications of our results for the single-parameter scaling hypothesis.
\end{abstract}

Pacs: 71.23.An, 72.15.Rn, 05.40.-a

\section{Introduction}

Although it was introduced more than fifty years ago~\cite{And58},
the tight-binding model named after Anderson is still widely studied
because it combines an elementary mathematical structure with non-trivial
physical features.
The simplicity of the definition notwithstanding, a complete analytical
understanding of the model is quite difficult to obtain, even in the
one-dimensional (1D) case which is the most amenable to analytical
treatment.
Some fundamental properties of the 1D Anderson model, however, have long
been known. In particular, it was rigorously proved that all eigenstates
are localised in the 1D Anderson model~\cite{Ish73}, unless the random
potential exhibits specific spatial correlations (see~\cite{Izr11} for
a comprehensive treatment of localisation in models with correlated
disorder).

The localisation length is the key physical parameter which determines the
spatial extension of the electronic states. A general formula for the
localisation length in the 1D Anderson model is still not known, but
expressions for the limit cases of strong and weak disorder have been
derived.
The weak-disorder case, in particular, can be studied using a perturbative
approach due to Thouless~\cite{Tho79}.
Thouless' method gives a formula for the inverse localisation length which
works very well for most energies inside the band of the disorder-free model.
This perturbative formula, however, has its flaws: in fact, shortly after
the publication of Ref.~\cite{Tho79}, numerical calculations showed that
Thouless' expression does not reproduce the correct value of the localisation
length at the band centre, i.e., for $E=0$~\cite{Czy81}. A short time later,
Kappus and Wegner were able to ascribe this discrepancy to a
resonance effect which leads to a breakdown of the standard perturbation
theory~\cite{Kap81}.
Almost at the same time, and without any explicit connection to the work
of Ref.~\cite{Czy81,Kap81}, an analytical expression of the localisation
length for $E=0$ was derived in~\cite{Sar81}.

A thorough study of the ``anomaly'' at the band centre was eventually
performed by Derrida and Gardner~\cite{Der84}.
These authors showed that a ``naive'' perturbative approach was bound
to fail not only at the band centre, but also for every other ``rational''
value of the energy, i.e., for $E = 2 \cos ( \pi r)$ with $r$ a rational
number. Such an approach, in fact, gives an expansion of the localisation
length which breaks down because some coefficients diverge.
To avoid this pitfall, Derrida and Gardner devised a specific
perturbative technique which allowed them to analyse the anomalous
behaviour of the localisation length in the neighbourhood of the band
centre.
In particular, Derrida and Gardner showed that the discrepancy between
the numerical value of the localisation length and the prediction of
Thouless' formula aroused because the leading term in the correct
expansion of the localisation length did not coincide with that obtained
in Thouless' expansion.
Derrida and Gardner also proved that a similar anomaly existed in
the neighbourhood of $E=\pm 2 \cos (\pi/3)= \pm 1$.
This anomaly, however, manifested itself only in the next-to-leading
term of Thouless' expansion and was therefore undetectable within the
second-order approximation.
Derrida and Gardner surmised the existence of anomalies of a similar
kind for all energies $E = 2 \cos ( \pi r)$ with $r$ a rational number.
This conjecture was later confirmed by the authors of Ref.~\cite{Bov88},
who focused their attention on the mathematical questions left open by
Derrida and Gardner.
Subsequently, in Ref.~\cite{Kus93,Gol94} a quasi-degenerate perturbation
theory was developed with the aim of obtaining a uniform asymptotic
expansion, in powers of the strength of the disorder, of the probability
distribution for the ratio of the wavefunction amplitudes at neighbouring
sites. Using this non-standard perturbative approach, the authors
of Ref.~\cite{Kus93,Gol94} re-obtained the results of Derrida and Gardner
for the anomalies of the 1D Anderson model at the band centre and the band
edge.

In the last few years the band-centre anomaly of the 1D Anderson model
has been considered from a new point of view, i.e., as a test case for
the validity of the single-parameter scaling (SPS) hypothesis (see,
e.g.,~\cite{Sch03,Hei04,Dey00} and references therein).
According to the SPS theory~\cite{Abr79,And80}, which is a cornerstone of our
present understanding of Anderson localisation, the probability distribution
of the conductance should depend only on a single free parameter.
Studies of the 1D Anderson model, however, have shown that this condition
does not hold when the energy lies close to the band
centre~\cite{Sch03,Hei04}.
This has led to a renewed interest for the anomalies of the 1D Anderson
model in the neighbourhood of the band centre~\cite{Sch03,Hei04,Dey00}.

The purpose of this paper is to provide a complete analysis of the
localisation length in the neighbourhood of the band centre by means of
the Hamiltonian map approach.
This formalism, introduced in~\cite{Izr95,Izr98}, has proved to be quite
an effective tool for the study of 1D and quasi-1D disordered models (see,
e.g.,~\cite{Izr11} and references therein).
The Hamiltonian map approach is based on the mathematical correspondence
between the 1D Anderson model and a classical parametric oscillator.
This analogy makes possible to associate quantum states of the Anderson
model to phase-space trajectories of the parametric oscillator. In this
way the phenomenon of quantum localisation can be understood in dynamical
terms as energetic instability of a stochastic oscillator~\cite{Tes00}.
Our first goal, therefore, is to use the Hamiltonian map approach to derive
in a mathematically simple and physically transparent way the results which
have been obtained by Derrida and Gardner and subsequent authors with
non-standard and intricate perturbative techniques.

The second goal of this paper is to provide a detailed analysis of the
transition from the anomalous behaviour of the model at the band centre
to the normal regime away from $E=0$.
We focus our study on two physical magnitudes: the localisation length
itself and the invariant distribution of the angle variable of the
parametric oscillator (whose dynamics we describe in terms of action-angle
variables).
We find that both magnitudes exhibit a gradual crossover from the
anomalous expressions they have at the band centre to the regular forms
which are found for higher values of the energy.
The slow progression of the invariant distribution towards a flat form
is important because it calls into question the validity of the
single-parameter scaling hypothesis. In fact, this hypothesis rests
on the so-called random phase approximation, which is equivalent to the
assumption that the invariant distribution of the angle variable be
uniform~\cite{Lam82}.
As for the localisation length, our results confirm the blurred character of
the transition which was recently pointed out in~\cite{Kan11}. The analysis
of Ref.~\cite{Kan11}, however, was based on purely numerical methods. Here
we follow a different approach: we first obtain an analytical formula for
the localisation length in the neighbourhood of the band centre and we
then check its validity within the band by comparing its predictions
with numerical results.

The paper is organised as follows. In Sec.~\ref{standard} we define
the model and we describe how the Hamiltonian map approach works in the
standard case. In Sec.~\ref{invdistr} we show how the random-phase
approximation fails in the neighbourhood of the band centre and we
derive the invariant distribution for the angle variable of the
Hamiltonian map. In Sec.~\ref{loclen} we use this result to obtain
a general formula for the inverse localisation length.
We draw our conclusions in Sec.~\ref{conclu}.

\section{Localisation length for non-resonant values of the energy}
\label{standard}

We consider the 1D Anderson model with weak disorder, defined by the
Schr\"{o}dinger equation
\begin{equation}
\psi_{n+1} + \psi_{n-1} + \varepsilon_{n} \psi_{n} = E \psi_{n} .
\label{andmod}
\end{equation}
Disorder is introduced in the model~(\ref{andmod}) via the site energies
$\varepsilon_{n}$ which are independent identically distributed random
variables with zero average $\langle \varepsilon_{n} \rangle = 0$ and
variance
\begin{equation}
\langle \varepsilon_{n}^{2} \rangle = \sigma^{2} \ll 1 .
\label{weakdis}
\end{equation}
Throughout this paper we will restrict our considerations to the
weak-disorder case defined by condition~(\ref{weakdis}). In this case
the first two moments of the random site energies $\varepsilon_{n}$ provide
sufficient information on the statistical properties of the
model~(\ref{andmod}). Our results, therefore, are valid regardless of
the specific form of the distribution of the site energies.

The electronic states of the Anderson model~(\ref{andmod}) can be analysed
in terms of the trajectories of a classical oscillator with
Hamiltonian
\begin{equation}
H = \frac{p^{2}}{2m} + \frac{1}{2}m\omega^{2} x^{2} \left[ 1 + \xi(t) \right]
\label{kickos}
\end{equation}
where $\xi(t)$ is a frequency noise constituted by a succession of delta
kicks of random strength $\xi_{n}$,
\begin{displaymath}
\xi(t) = \sum_{n = -\infty}^{\infty} \xi_{n} \delta \left( t - n T \right).
\end{displaymath}
By integrating the dynamical equation of the kicked oscillator over the
time interval between two kicks, one obtains the Hamiltonian map
\begin{equation}
\begin{array}{ccl}
x_{n+1} & = & \displaystyle
\left[ \cos \left( \omega T \right) - \omega \xi_{n} \sin \left(
\omega T \right) \right] x_{n} + \frac{1}{m\omega} \sin \left( \omega T
\right) p_{n} \\
p_{n+1} & = & \displaystyle
- m \omega \left[ \sin \left( \omega T \right) + \omega \xi_{n}
\cos \left( \omega T \right) \right] x_{n} +
\cos \left( \omega T \right) p_{n} .
\end{array}
\label{hammap}
\end{equation}
Here we have used the symbols $x_{n} = x(t_{n}^{-})$ and $p_{n} =
p(t_{n}^{-})$ to represent the position and the momentum of the
oscillator immediately before the $n\mbox{-th}$ kick.
After eliminating the momenta from the map~(\ref{hammap}), one obtains the
equation
\begin{equation}
x_{n+1} + x_{n-1} + \omega \xi_{n} \sin \left( \omega T \right)
x_{n} = 2 \cos \left( \omega T \right) x_{n}
\label{nomomenta}
\end{equation}
which has the same form of the Schr\"{o}dinger equation~(\ref{andmod})
for the Anderson model provided that the following identities hold
\begin{equation}
E = 2 \cos \left( \omega T \right)
\label{energy}
\end{equation}
and
\begin{equation}
\varepsilon_{n} = \omega \xi_{n} \sin \left( \omega T \right) .
\label{epsxi}
\end{equation}
Condition~(\ref{energy}) corresponds to the dispersion relation for
the Anderson model, if one makes the identification $\omega T = \mu a$
linking the frequency $\omega$ of the kicked oscillator and the period $T$
of the kicks with the wavenumber $\mu$ of the electronic states and
the lattice step $a$ of the model~(\ref{andmod}).
Eq.~(\ref{epsxi}), on the other hand, allows one to link the spatial
disorder in the Anderson model~(\ref{andmod}) with the frequency noise
of the parametric oscillator~(\ref{kickos}).

The dynamical analysis of the kicked oscillator~(\ref{kickos}) becomes
simpler if the Hamiltonian map~(\ref{hammap}) is written in action-angle
variables introduced through the canonical transformation
\begin{displaymath}
\begin{array}{ccl}
x_{n} & = & \displaystyle
\sqrt{\frac{2J_{n}}{m\omega}} \sin \theta_{n} \\
p_{n} & = & \displaystyle
\sqrt{2 m \omega J_{n}} \cos \theta_{n} .
\end{array}
\end{displaymath}
One thus obtains
\begin{equation}
\begin{array}{ccl}
\sin \theta_{n+1} & = & \displaystyle
\frac{1}{D_{n}} \left[ \sin \left( \theta_{n} + \omega T \right)
- \omega \xi_{n} \sin \theta_{n} 
\sin \left( \omega T \right) \right] \\
\cos \theta_{n+1} & = & \displaystyle
\frac{1}{D_{n}} \left[ \cos \left( \theta_{n} + \omega T \right)
- \omega \xi_{n} \sin \theta_{n} 
\cos \left( \omega T \right) \right] \\
\end{array}
\label{hamil1}
\end{equation}
with
\begin{equation}
D_{n}^{2} = \frac{J_{n+1}}{J_{n}} = 1 - 2 \omega \xi_{n} \sin \theta_{n}
\cos \theta_{n} + \omega^{2} \xi_{n}^{2} \sin^{2} \theta_{n}.
\label{hamil2}
\end{equation}
Making use of the weak-disorder condition~(\ref{weakdis}), the map
expressed by Eqs.~(\ref{hamil1}) and~(\ref{hamil2}) can be written
in the more compact form
\begin{eqnarray}
\begin{array}{c}
\theta_{n+1} \\
\\
\end{array}
&\begin{array}{c} = \\ + \\ \end{array} &
\begin{array}{l} 
\theta_{n} + \omega T + \omega \xi_{n} \sin^{2} \theta_{n}
+ (\omega \xi_{n})^{2} \sin^{3} \theta_{n} \cos \theta_{n} \\
O\left( \sigma^{3} \right) \pmod{2 \pi}
\end{array}
\label{map} \\
J_{n+1} & = & J_{n} \left( 1 - 2 \omega \xi_{n} \sin \theta_{n} \cos \theta_{n}
+ \omega^{2} \xi_{n}^{2} \sin^{2} \theta_{n} \right) .
\label{jmap}
\end{eqnarray}
In Eq.~(\ref{map}) we have used the Landau symbol $O(\sigma^{3})$ to denote
neglected terms which, in the limit $\sigma \to 0$, vanish like $\sigma^{3} =
(\langle \varepsilon_{n}^{2} \rangle)^{3/2}$ or faster (see,
e.g.,~\cite{Har60}).

The inverse localisation length (or Lyapunov exponent) is defined as
\begin{displaymath}
\lambda = \lim_{N \rightarrow \infty} \frac{1}{N} \sum_{n=1}^{N}
\log \left| \frac{\psi_{n}}{\psi_{n-1}} \right| =
\lim_{N \rightarrow \infty} \frac{1}{N} \sum_{n=1}^{N}
\log \left| \frac{x_{n}}{x_{n-1}} \right| .
\end{displaymath}
In terms of the action-angle variables one can write the previous expression
in the form
\begin{displaymath}
\lambda = \lim_{N \rightarrow \infty} \frac{1}{N} \sum_{n=1}^{N}
\log D_{n-1} + \lim_{N \rightarrow \infty} \frac{1}{N}
\log \left| \frac{\sin \theta_{N}}{\sin \theta_{0}} \right| .
\end{displaymath}
Except that at the band edge (where the angular variable tends to assume
the values $0$ and $\pi$), the second term in the right-hand side (rhs)
of this identity vanishes; one is therefore left with
\begin{equation}
\lambda = \lim_{N \rightarrow \infty} \frac{1}{N} \sum_{n=1}^{N}
\log D_{n-1} = \langle \log D_{n} \rangle .
\label{lambda1}
\end{equation}
For weak and uncorrelated disorder one can expand the logarithm in the
rhs of Eq.~(\ref{lambda1}) and factorise the disorder-angle
correlators. Within the second-order approximation one thus obtains
\begin{equation}
\lambda = \frac{\omega^{2}}{8} \langle \xi_{n}^{2} \rangle
\left[ 1 - 2 \langle \cos \left( 2 \theta_{n} \right) \rangle +
\langle \cos \left( 4 \theta_{n} \right) \rangle \right]
+ O\left( \sigma^{3} \right) .
\label{lyap1}
\end{equation}
To proceed further, one must evaluate the averages of the trigonometric
functions in the rhs of Eq.~(\ref{lyap1}); it is therefore necessary
to determine the invariant distribution $\rho (\theta)$ for the angular
map~(\ref{map}).

Comparing Eqs.~(\ref{map}) and~(\ref{jmap}), it is easy to see that
the angle variable has a faster dynamics than the action variable and
that a limited number of map iterations suffices to make the $\theta$
variable sweep the whole $[0,2\pi]$ interval.
One can therefore expect that the distribution for the angle variable
should quickly reach a uniform invariant form,
\begin{equation}
\rho(\theta) = \frac{1}{2 \pi} .
\label{unidistr}
\end{equation}
We remark that the assumption that the invariant distribution has the flat
form~(\ref{unidistr}) corresponds to the so-called random phase
approximation~\cite{Sch03,Lam82}.
This approximation assumes that, for a random sample of length $L$, the
phases of the reflection and transmission amplitudes are uniformly
distributed over $[0,2\pi]$ when the length of the sample is much larger
than the localisation length, $L \gg 1/\lambda$.
The random phase approximation is usually invoked in the SPS theory
of localisation~\cite{Sch03,And80}; when condition~(\ref{unidistr})
is not met, therefore, one may expect the SPS theory to fail.

Besides being a key ingredient in the derivation of the SPS theory,
condition~(\ref{unidistr}) is also required to obtain Thouless' formula.
In fact, Thouless' expression for the localisation length can be easily
recovered from Eq.~(\ref{lyap1}) by computing the averages in the rhs of
this equation with the invariant distribution~(\ref{unidistr}).
In this way one arrives at the formula
\begin{equation}
\lambda = \frac{1}{8} \omega^{2} \langle \xi_{n}^{2} \rangle =
\frac{\sigma^{2}}{8 \left( 1 - E^{2}/4 \right)} .
\label{thouless}
\end{equation}

The analysis of the map~(\ref{map}) shows that the
assumption~(\ref{unidistr}) is usually correct; however, it fails
whenever the parameter $\omega T$ is a rational multiple of $\pi$.
For $\omega T = \pi/q$, in fact, the angular map~(\ref{map})
has periodic orbits of period $2q$ in the absence of disorder.
When weak disorder is introduced, the resulting orbits linger around
the unperturbed periodic orbits; as a result, the invariant distribution
ceases to be flat.
The modulation of the invariant measure is the origin of the failure
of Thouless' formula~(\ref{thouless}) at the band edge, i.e.,
for $\omega T = 0$ or $\omega T = \pm \pi$, and at the band centre,
i.e., for $\omega T = \pi/2$.
In general terms, the invariant distribution acquires a modulation
for all the ``rational'' values of the energy $E = 2 \cos (\pi p/q)$;
however, within the limits of the second-order approximation, the
localisation length differs from the standard value predicted by
Eq.~(\ref{thouless}) only for $|q| \le 2$, i.e., at the
band edge and at the band centre. Higher values of $q$, in fact,
produce a high-frequency modulation of the invariant distribution
which does not affect the averages of the low-order harmonics in the
rhs of Eq.~(\ref{lyap1}).

The anomalous behaviour in the neighbourhood of the band edge was
thoroughly analysed with the Hamiltonian map approach in~\cite{Izr98}.
We now turn our attention to the neighbourhood of the band centre,
devoting the next section to the derivation of the invariant distribution
in this region.

\section{The invariant distribution}
\label{invdistr}

We consider the case in which
\begin{equation}
\omega T = \frac{\pi}{2} + \delta
\label{nbc}
\end{equation}
with $\delta \rightarrow 0$.
Taking into account the dispersion relation~(\ref{energy}), for $\delta \to
0$ Eq.~(\ref{nbc}) implies that
\begin{displaymath}
E = -2 \sin \delta \simeq -2 \delta,
\end{displaymath}
i.e., the energy lies in a neighbourhood of the band centre.
From Eq.~(\ref{map}) it is easy to see that, when condition~(\ref{nbc}) is
met, the unperturbed angular map has almost-periodic orbits of period 4.
This implies that $\theta_{n+4}$ must lie close to $\theta_{n}$ even when
disorder is present. We can compute the difference $\theta_{n+4} - \theta_{n}$
by considering the fourth iterate of the angular map~(\ref{map}).
Neglecting terms of the form $\xi_{n} \delta$, which are of order
$O(\sigma \delta)$, and making use of the statistical independence of the
variables $\xi_{n}$, one can write the fourth iterate of the angular
map~(\ref{map}) in the form
\begin{equation}
\begin{array}{ccl}
\theta_{n+4} & \simeq & \displaystyle
\theta_{n} + 4 \delta + \frac{\omega}{2} \left( \xi_{n} +
\xi_{n+2} \right) \left[ 1 - \cos \left( 2 \theta_{n} \right) \right] \\
& + & \displaystyle
\frac{\omega}{2} \left( \xi_{n+1} + \xi_{n+3} \right) \left[ 1 +
\cos  \left( 2 \theta_{n} \right) \right]
- \frac{\sigma^{2}}{2} \sin \left( 4 \theta_{n} \right) \pmod{2 \pi}. \\
\end{array}
\label{fourth}
\end{equation}
The sign of approximate identity in Eq.~(\ref{fourth}) is due to the
fact that the rhs is a truncated expansion in both the $\sigma$ and
$\delta$ variables in which we omit $O \left( \sigma^{3} \right)$ terms
of order higher than the second in the disorder strength, second-order
terms $O \left( \delta^{2} \right)$ in the energy shift $\delta$, and
cross-terms $O(\sigma \delta)$.
Note that, in the limit case $\delta = 0$, Eq.~(\ref{fourth}) reduces to the
form used in~\cite{Izr98} to analyse the exact band-centre case.
By going to the continuum limit one can replace the map~(\ref{fourth}) with
the It\^{o} stochastic differential equation
\begin{equation}
\begin{array}{ccl}
\mathrm{d} \theta & = & \displaystyle
\left[ 4 \delta + \frac{\sigma^{2}}{2}
\sin \left( 4 \theta \right) \right] \mathrm{d}t 
 + \sqrt{\frac{\sigma^{2}}{2}}
\left[ 1 - \cos \left( 2 \theta\right) \right] \mathrm{d} W_{1}\\
& + & \displaystyle
\sqrt{\frac{\sigma^{2}}{2}}
\left[ 1 + \cos \left( 2 \theta \right) \right] \mathrm{d} W_{2}\\
\end{array}
\label{sde}
\end{equation}
where $W_{1}(t)$ and $W_{2}(t)$ are independent Wiener processes.
The It\^{o} stochastic equation~(\ref{sde}), together with an initial
condition $\theta(t_{0}) = \theta_{0}$, completely determines the stochastic
process $\theta(t)$.

As is well known~\cite{Gar04}, the $\theta(t)$ process can be specified
also in terms of the conditional probability distribution
\begin{displaymath}
p = p \left( \theta, t | \theta_{0}, t_{0} \right)
\end{displaymath}
which is obtained by solving the associated Fokker-Planck equation
\begin{equation}
\frac{\partial p}{\partial t} =
\frac{\sigma^{2}}{4} \left\{ \frac{\partial}{\partial \theta}
\left[ \left( -16 \frac{\delta}{\sigma^{2}} +
2 \sin \left( 4 \theta \right) \right) p \right] +
\frac{\partial^{2}}{\partial \theta^{2}} \left[ \left( 3 + \cos
\left( 4 \theta \right) \right) p \right]
\right\}
\label{fp_raw}
\end{equation}
with the initial condition $p \left( \theta, t_{0} | \theta_{0}, t_{0} \right)
= \delta (\theta - \theta_{0})$.

It is convenient to introduce the parameter
\begin{equation}
\varkappa = - \frac{2 \delta}{\sigma^{2}} \simeq \frac{E}{\sigma^{2}}
\label{kappa}
\end{equation}
which measures the distance of the energy from the band centre on a
scale defined by the strength of the disorder.
From a mathematical point of view, the parameter~(\ref{kappa}) represents
a ratio which is held fixed while the double limit
\begin{eqnarray*}
E \rightarrow 0 & \mbox{ and } &  \sigma^{2} \rightarrow 0
\end{eqnarray*}
is taken.
Inserting the new parameter in Eq.~(\ref{fp_raw}) and rescaling time
according to the identity
\begin{displaymath}
\tau = \frac{\sigma^{2}}{4} t,
\end{displaymath}
one can cast the Fokker-Planck equation~(\ref{fp_raw}) in the form
\begin{equation}
\frac{\partial p}{\partial \tau} =
\frac{\partial}{\partial \theta} \left\{ \left[ 8 \varkappa +
2 \sin (4 \theta) \right] p + \frac{\partial}{\partial \theta}
\left[ \left(3 + \cos (4 \theta) \right) p \right] \right\} .
\label{fp}
\end{equation}

The stationary solution of the Fokker-Planck Eq.~(\ref{fp}) is the
invariant distribution $\rho(\theta)$ we are interested into.
Therefore we must find the solution of the stationary equation
\begin{equation}
\frac{d}{d \theta} \left\{ \left[ 8 \varkappa +
2 \sin \left( 4 \theta \right) \right] \rho \left( \theta \right) +
\frac{d}{d \theta} \left[ \left( 3 + \cos \left( 4 \theta \right) \right)
\rho \left( \theta \right) \right] \right\} = 0
\label{sfp}
\end{equation}
which satisfies the conditions of periodicity
\begin{equation}
\rho \left( \theta + 2 \pi \right) =  \rho \left( \theta \right)
\label{periodic}
\end{equation}
and normalisation
\begin{displaymath}
\int_{0}^{2 \pi} \rho \left( \theta \right) d \theta = 1 .
\end{displaymath}

Integrating once Eq.~(\ref{sfp}) one obtains the first-order ordinary
differential equation
\begin{displaymath}
\frac{d \rho}{d \theta} = \frac{2 \sin \left(4 \theta \right) -
8\varkappa} {3 + \cos \left( 4 \theta \right)} \rho +
\frac{C}{3 + \cos \left( 4 \theta \right)}
\end{displaymath}
with $C$ being an integration constant.
The general solution of this equation has the form
\begin{equation}
\rho(\theta) = \frac{2e^{-8 \varkappa F(\theta)}}{\sqrt{3 + \cos(4\theta)}}
\left[ \rho(0) + \frac{C}{2} \int_{0}^{\theta}
\frac{e^{8 \varkappa F(\phi)}}{\sqrt{3 + \cos \left( 4 \phi \right)}}
\mathrm{d}\phi \right]
\label{rho_raw}
\end{equation}
with
\begin{equation}
\begin{array}{cclcc}
F (\theta) & = & \displaystyle \int_{0}^{\theta}
\frac{1}{3 + \cos \left( 4\phi \right)} \mathrm{d}\phi & & \\
& = & \displaystyle
\frac{1}{4\sqrt{2}} \left\{ \pi n + \arctan \left[ \frac{1}{\sqrt{2}}
\tan (2 \theta) \right] \right\} & \mbox{ for } &
\displaystyle
n \frac{\pi}{2} \le \theta \le (n+1) \frac{\pi}{2} .
\end{array}
\label{auxfun}
\end{equation}
In Eq.~(\ref{auxfun}) $n =0, \pm 1, \pm 2, \ldots$ and arctan denotes
the inverse of the tangent function having the interval $[0,\pi]$ as
domain.
The solution~(\ref{rho_raw}) contains the constants $C$ and $\rho(0)$
which must be determined from the periodicity and normalisation
conditions.
The periodicity condition~(\ref{periodic}) implies that
\begin{displaymath}
C = 2 \rho(0) \left( e^{4 \sqrt{2} \pi \varkappa} - 1\right) \left/
\int_{0}^{2 \pi} \frac{e^{8 \varkappa F(\phi)}}{\sqrt{3 + \cos(4\phi)}}
\mathrm{d} \phi \right. .
\end{displaymath}
Substituting this identity in Eq.~(\ref{rho_raw}) and imposing the
normalisation condition, one eventually obtains that the invariant
distribution has the form
\begin{equation}
\rho(\theta) = \frac{1}{N(\varkappa)}
\frac{e^{-8 \varkappa F(\theta)}}{\sqrt{3 + \cos(4\theta)}}
\left[e^{4 \sqrt{2} \pi \varkappa}
\int_{0}^{\theta} \frac{e^{8 \varkappa F(\phi)}}{\sqrt{3 + \cos(4\phi)}}
\mathrm{d}\phi +
\int_{\theta}^{2 \pi} \frac{e^{8 \varkappa F(\phi)}}{\sqrt{3 + \cos(4\phi)}}
\mathrm{d}\phi \right]
\label{invdis}
\end{equation}
where the normalisation factor $N(\varkappa)$ is equal to
\begin{equation}
N(\varkappa) = \int_{0}^{2 \pi}
\frac{e^{-8 \varkappa F(\theta)}}{\sqrt{3 + \cos(4\theta)}}
\left[e^{4 \sqrt{2} \pi \varkappa}
\int_{0}^{\theta} \frac{e^{8 \varkappa F(\phi)}}{\sqrt{3 + \cos(4\phi)}}
\mathrm{d}\phi +
\int_{\theta}^{2 \pi} \frac{e^{8 \varkappa F(\phi)}}{\sqrt{3 + \cos(4\phi)}}
\mathrm{d}\phi \right] \mathrm{d} \theta .
\label{norm}
\end{equation}
To the best of our knowledge, the general expression~(\ref{invdis}) of
the invariant distribution has not been discussed before.
Ref.~\cite{Gol94} contains equations which, if combined, are equivalent
to Eq.~(\ref{invdis}) but the authors of Ref.~\cite{Gol94} worked out
the explicit form of the invariant distribution only for the limit cases
of $|\varkappa| \ll 1$ and $|\varkappa| \gg 1$.
The limit forms of the invariant distribution are much easier to use (see
below), but in this paper we prefer to focus our attention on the general
expression~(\ref{invdis}), because it is the only distribution that
allows one to analyse the behaviour of the localisation length over
the whole neighbourhood of the band centre.

An important property of the invariant distribution~(\ref{invdis}) is
that it is $\pi/2$-periodic, i.e.,
\begin{equation}
\rho \left( \theta + \frac{\pi}{2} \right) = \rho \left( \theta \right) .
\label{periodicity}
\end{equation}
In physical terms the periodicity condition~(\ref{periodicity})
should be seen as an expected manifestation of the fact that
unperturbed orbits of the map~(\ref{map}) have period 4 for $E=0$.
From a mathematical point of view, Eq.~(\ref{periodicity}) can be
obtained by first noting that the definition~(\ref{auxfun}) implies that
\begin{equation}
F \left( \theta + \frac{\pi}{2} \right) =
F \left( \theta \right) + \frac{\pi}{4 \sqrt{2}} .
\label{fcond}
\end{equation}
Keeping in mind Eqs.~(\ref{auxfun}) and~(\ref{fcond}), one can analyse
the behaviour of the integral terms in the representation~(\ref{invdis})
of the invariant distribution. It is easy to show that
\begin{equation}
\begin{array}{ccl}
\displaystyle
\int_{0}^{\theta + \pi/2} \frac{e^{8 \varkappa F(\phi)}}{\sqrt{3 + \cos (4 \phi)}}
\mathrm{d} \phi & = & \displaystyle
\int_{0}^{\pi/2} \frac{e^{8 \varkappa F(\phi)}}{\sqrt{3 + \cos (4 \phi)}}
\mathrm{d} \phi \\
& + & \displaystyle
e^{\sqrt{2} \pi \varkappa}
\int_{0}^{\theta} \frac{e^{8 \varkappa F(\phi)}}{\sqrt{3 + \cos (4 \phi)}}
\mathrm{d} \phi
\end{array}
\label{integ1}
\end{equation}
and
\begin{equation}
\begin{array}{ccl}
\displaystyle
\int_{\theta + \pi/2}^{2 \pi} \frac{e^{8 \varkappa F(\phi)}}{\sqrt{3 + \cos(4 \phi)}}
\mathrm{d} \phi & = & \displaystyle
e^{\sqrt{2}\pi \varkappa} \int_{\theta}^{2 \pi} 
\frac{e^{8 \varkappa F(\phi)}}{\sqrt{3 + \cos (4 \phi)}} \mathrm{d} \phi \\
& - & \displaystyle
e^{4\sqrt{2}\pi \varkappa} \int_{0}^{\pi/2} 
\frac{e^{8 \varkappa F(\phi)}}{\sqrt{3 + \cos (4 \phi)}} \mathrm{d} \phi .
\end{array}
\label{integ2}
\end{equation}
Making use of the identities~(\ref{fcond}),~(\ref{integ1}),
and~(\ref{integ2}), one can evaluate $\rho( \theta + \pi/2)$ from
Eq.~(\ref{invdis}) and thus obtain the periodicity
condition~(\ref{periodicity}).

\begin{figure}[t]
\begin{center}
\includegraphics[width=5in,height=3.5in]{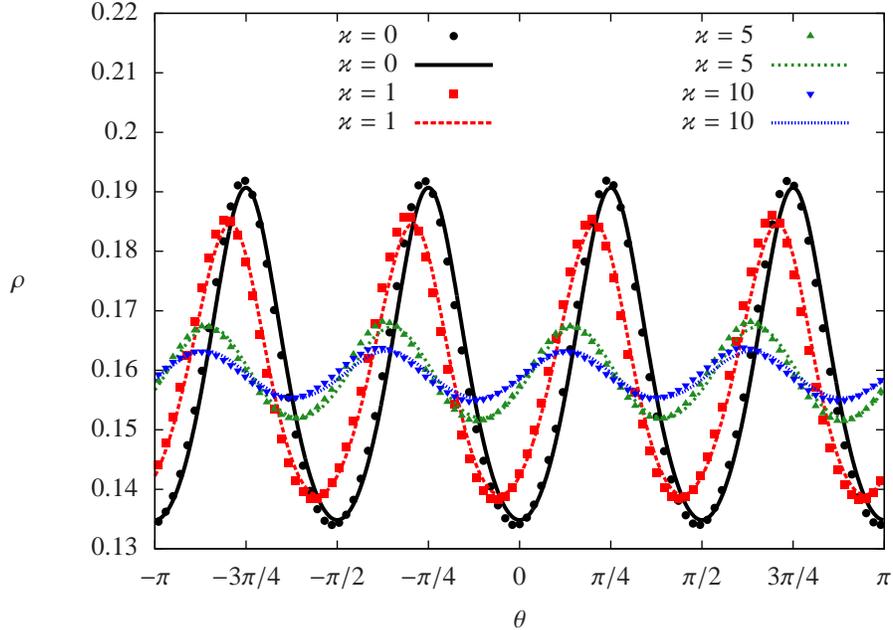}
\caption{\label{invdis_fig}
(Colour online) Invariant distribution $\rho(\theta)$ for various values of
$\varkappa$.
The points represent the numerically computed invariant distribution,
while the lines correspond to Eq.~(\ref{invdis}).}
\end{center}
\end{figure}

\begin{figure}[t]
\begin{center}
\includegraphics[width=5in,height=3.5in]{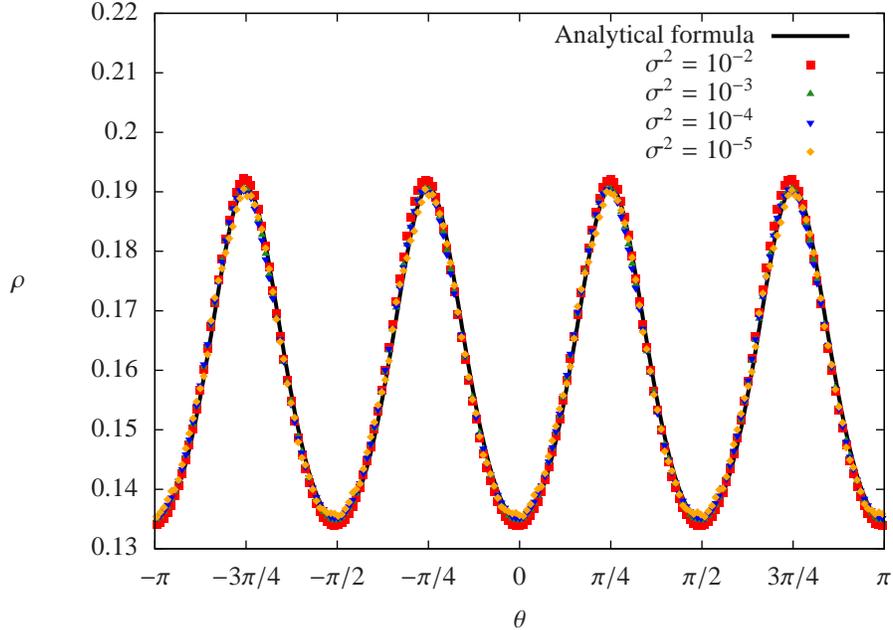}
\caption{\label{collapse}
(Colour online) Invariant distribution $\rho(\theta)$ at the band centre
($E=0$) for various values of $\sigma^{2}$.
The points represent the numerically computed invariant distribution,
while the solid line corresponds to Eq.~(\ref{invdis}).}
\end{center}
\end{figure}

The validity of the analytical expression~(\ref{invdis}) is corroborated
by numerical computations, as can be seen from Figs.~\ref{invdis_fig}
and~\ref{collapse}.
In Fig.~\ref{invdis_fig} the theoretical distribution~(\ref{invdis}) is
plotted together with the distributions which were obtained numerically
for several values of $\varkappa$.
As can be clearly seen, the invariant distribution~(\ref{invdis})
matches very well the numerical data.
Fig.~\ref{invdis_fig} confirms that the invariant distribution is
not uniform in the neighbourhood of the band centre. The amplitude of
the modulation is largest at the exact band centre, i.e., for
$\varkappa = 0$, and gradually decreases as as the energy moves away from the
band centre (i.e., for increasing values of $|\varkappa|$).

The numerical data displayed in Fig.~\ref{invdis_fig} were obtained
for disorder strength $\sigma^{2} = 0.01$, but the results are the same
for different values of $\sigma^{2}$.
This is made evident by Fig.~\ref{collapse}, which represents the
invariant distribution of the map~(\ref{map}) at the band centre ($E=0$)
computed for different values of the disorder strength.
The good data collapse on the line corresponding to the theoretical
form~(\ref{invdis}) confirms another feature of Eq.~(\ref{invdis}), i.e.,
that in the neighbourhood of the band centre the invariant distribution
is determined by the parameter $\varkappa$ and does not depend on the
disorder strength.
The data in Fig.~\ref{collapse} were obtained for $\varkappa=0$ but the
same collapse occurs also for non-vanishing values of $\varkappa$.

As mentioned before, the general expression~(\ref{invdis}) for the
invariant distribution reduces to simpler forms in the limit cases
$|\varkappa| \ll 1$ and $|\varkappa| \gg 1$, which are the only ones
considered in the literature~\cite{Der84,Gol94}. We devote the rest
of this Section to the analysis of these particular cases.

\subsection{The invariant distribution for $|\varkappa| \ll 1$}

In the limit case $|\varkappa| \ll 1$ one can expand both the
expression~(\ref{invdis}) and the normalisation factor~(\ref{norm})
in powers of $\varkappa$. Carrying out the calculations, one
eventually obtains that the limit form of the invariant distribution is
\begin{equation}
\rho(\theta) = \frac{2 \rho(0)}{\sqrt{3 + \cos (4 \theta)}}
\left\{ 1 + \varkappa \left[ -8 F(\theta) + 8 \sqrt{\pi}
\frac{\Gamma\left(\frac{3}{4}\right)}{\Gamma \left( \frac{1}{4} \right)}
\int_{0}^{\theta} \frac{1}{\sqrt{3 + \cos(4\phi)}} \mathrm{d}\phi \right]
+ O(\varkappa^{2}) \right\}
\label{rhosmall}
\end{equation}
with the constant factor $2 \rho(0)$ given by
\begin{displaymath}
\begin{array}{ccl}
2 \rho(0) & = & \displaystyle
\sqrt{\frac{2}{\pi}}
\frac{\Gamma\left(\frac{3}{4}\right)}{\Gamma \left( \frac{1}{4} \right)}
\left\{ 1 + \varkappa \left[ 8 \sqrt{\frac{2}{\pi}}
\frac{\Gamma\left(\frac{3}{4}\right)}{\Gamma \left( \frac{1}{4} \right)}
\int_{0}^{2 \pi} \frac{F(\phi)}{\sqrt{3 + \cos(4\phi)}} \mathrm{d} \phi
\right. \right.\\
& - & \displaystyle \left. \left.
8 \sqrt{\frac{2}{\pi}} 
\left[\frac{\Gamma\left(\frac{3}{4}\right)}{\Gamma \left( \frac{1}{4} \right)}
\right]^{2}
\int_{0}^{2 \pi} \frac{1}{\sqrt{3 + \cos (4\theta)}} \int_{0}^{\theta}
\frac{1}{\sqrt{3 + \cos(4\phi)}} \mathrm{d}\phi \mathrm{d}\theta \right]
\right\} + O(\varkappa^{2}).
\end{array}
\end{displaymath}
Eq.~(\ref{rhosmall}) coincides with the corresponding result first obtained
by Derrida and Gardner~\cite{Der84}; the apparent difference is due to
the election of these authors of a complex representation for
the function~(\ref{auxfun}).

At the band centre, i.e., for $\varkappa = 0$, expression~(\ref{rhosmall})
reduces to the form
\begin{displaymath}
\rho(\theta) = \sqrt{\frac{2}{\pi}} 
\frac{\Gamma\left(\frac{3}{4}\right)}{\Gamma \left( \frac{1}{4} \right)}
\frac{1}{\sqrt{3 + \cos(4 \theta)}} =
\frac{1}{2 \mathbf{K} \left(\frac{1}{\sqrt{2}}\right)
\sqrt{3 + \cos(4 \theta)}} .
\end{displaymath}
In the previous equation the symbol $\mathbf{K}(k)$ stands for the complete
elliptic integral of the first kind,
\begin{displaymath}
\mathbf{K}(k) = \int_{0}^{\pi/2} \frac{d \phi}{\sqrt{1 - k^{2}
\sin^{2} \phi}}
\end{displaymath}
and we used the identity
\begin{displaymath}
\mathbf{K}\left( \frac{1}{\sqrt{2}} \right) = \sqrt{\frac{\pi}{8}}
\frac{\Gamma\left(\frac{1}{4}\right)}{\Gamma \left( \frac{3}{4} \right)} .
\end{displaymath}

\subsection{The invariant distribution for $|\varkappa| \gg 1$}
\label{kappalarge}

We now consider the case $|\varkappa| \gg 1$. In this limit, the
asymptotic expansion of $\rho(\theta)$ is obtained with less effort
taking as a starting point the differential equation~(\ref{sfp}), rather
than the general form~(\ref{invdis}) of the invariant distribution.
After writing Eq.~(\ref{sfp}) as
\begin{equation}
\frac{d}{d \theta} \left[ \rho(\theta) - \frac{1}{4 \varkappa}
\sin (4 \theta) \rho(\theta) + \frac{1}{8 \varkappa} \left(
3 + \cos (4 \theta) \right) \frac{d \rho}{d \theta} \right] = 0 ,
\label{sfp2}
\end{equation}
one can expand the solution in powers of $1/\varkappa$,
\begin{equation}
\rho(\theta) = \rho^{(0)}(\theta) + \frac{1}{\varkappa} \rho^{(1)}(\theta)
+ \frac{1}{\varkappa^{2}} \rho^{(2)}(\theta) + \ldots
\label{rho_exp}
\end{equation}
Substituting the expansion~(\ref{rho_exp}) in Eq.~(\ref{sfp2}) one
obtains the hierarchy of equations
\begin{equation}
\frac{d \rho^{(0)}}{d \theta} = 0
\label{zeroth}
\end{equation}
and
\begin{equation}
\frac{d}{d \theta} \left[ \rho^{(n)} - \frac{1}{4} \sin (4 \theta) \rho^{(n-1)} +
\frac{1}{8} \left(3 + \cos(4\theta) \right) \frac{d \rho^{(n-1)}}{d \theta}
\right] = 0 
\label{nth}
\end{equation}
with $n = 1, 2, \ldots$
For the expansion~(\ref{rho_exp}) to satisfy the condition of periodicity
and normalisation, the solution of Eq.~(\ref{zeroth}) must be $2\pi$-periodic
and normalised, while the solutions of the higher-order equations~(\ref{nth})
must be $2\pi$-periodic and satisfy the conditions
\begin{displaymath}
\int_{0}^{2 \pi} \rho^{(n)}(\theta) \mathrm{d} \theta = 0 .
\end{displaymath}
In this way one can easily obtain that the behaviour of the invariant
distribution for $\varkappa \gg 1$ is
\begin{equation}
\rho(\theta) = \frac{1}{2\pi} + \frac{1}{\varkappa} \frac{1}{8\pi}
\sin (4\theta) - \frac{1}{\varkappa^{2}} \frac{3}{16 \pi}
\left[ \cos(4\theta) + \frac{1}{4} \cos(8 \theta) \right] +
O \left( \frac{1}{\varkappa^{3}} \right) .
\label{rholarge}
\end{equation}
Eq.~(\ref{rholarge}) coincides with the result previously obtained
in~\cite{Der84}. Note that as $|\varkappa|$ increases, i.e., as the energy
moves away from the band centre, the invariant distribution tends to
recover its flat form. The uniform limit, however, is reached with
power-law convergence: this implies that the transition from anomalous
to regular behaviour is smeared out over a wide energy range.

\section{The localisation length}
\label{loclen}

Having obtained the invariant distribution~(\ref{invdis}), one can
compute the rhs of expression~(\ref{lyap1}). After observing that
the average of $\cos(2 \theta)$ vanishes because the invariant distribution
has period $\pi/2$, one obtains that the inverse localisation length
is equal to
\begin{equation}
\begin{array}{ccl}
\lambda & \simeq & \displaystyle
\frac{\sigma^{2}}{8 \left( 1 + \frac{E^{2}}{4} \right)}
\left\{ 1 + \frac{1}{N(\varkappa)}
\int_{0}^{2\pi} \frac{\cos(4\theta)}{\sqrt{3 + \cos(4\theta)}}
e^{-8 \varkappa F(\theta)} \right. \\
& \times & \displaystyle \left.
\left[e^{4 \sqrt{2} \pi \varkappa} \int_{0}^{\theta}
\frac{e^{8 \varkappa F(\phi)}}{\sqrt{3 + \cos(4\phi)}} \mathrm{d}\phi +
\int_{\theta}^{2 \pi} \frac{e^{8 \varkappa F(\phi)}}{\sqrt{3 + \cos(4\phi)}}
\mathrm{d}\phi \right] \mathrm{d} \theta \right\} .
\end{array}
\label{inter}
\end{equation}
The general expression~(\ref{inter}) is the central result of this
paper. Its absence from previous works on the band-centre anomaly is
probably explained by the fact that in the literature only the limit
cases $|\varkappa| \ll 1$ and $|\varkappa| \gg 1$ have received attention
so far.
We discuss these limit cases below; we would like to emphasise, however,
that only the general expression~(\ref{inter}) makes possible to describe
the behaviour of the localisation length over the whole range of the
parameter~$\varkappa$.

A remark is in order here. Because the invariant distribution~(\ref{invdis})
was obtained under the assumption $E \simeq -2\delta \to 0$, in the
energy range where Eq.~(\ref{inter}) is {\em rigorously} valid, it is
equivalent to the following form
\begin{equation}
\begin{array}{ccl}
\lambda & \simeq & \displaystyle
\frac{\sigma^{2}}{8}
\left\{ 1 + \frac{1}{N(\varkappa)}
\int_{0}^{2\pi} \frac{\cos(4\theta)}{\sqrt{3 + \cos(4\theta)}}
e^{-8 \varkappa F(\theta)} \right. \\
& \times & \displaystyle \left.
\left[e^{4 \sqrt{2} \pi \varkappa} \int_{0}^{\theta}
\frac{e^{8 \varkappa F(\phi)}}{\sqrt{3 + \cos(4\phi)}} \mathrm{d}\phi +
\int_{\theta}^{2 \pi} \frac{e^{8 \varkappa F(\phi)}}{\sqrt{3 + \cos(4\phi)}}
\mathrm{d}\phi \right] \mathrm{d} \theta \right\} .
\end{array}
\label{general}
\end{equation}
which is obtained by setting $E=0$ in Eq.~(\ref{inter}).
However, as discussed below, we have found that Eq.~(\ref{inter}) has
a validity of its own because it provides a very effective interpolation
between formula~(\ref{general}), which is applicable in a neighbourhood of
the band centre, and Thouless' expression~(\ref{thouless}) which can be
used in the rest of the band (with the exception of the band edges).

\begin{figure}[t]
\begin{center}
\includegraphics[width=5in,height=3.5in]{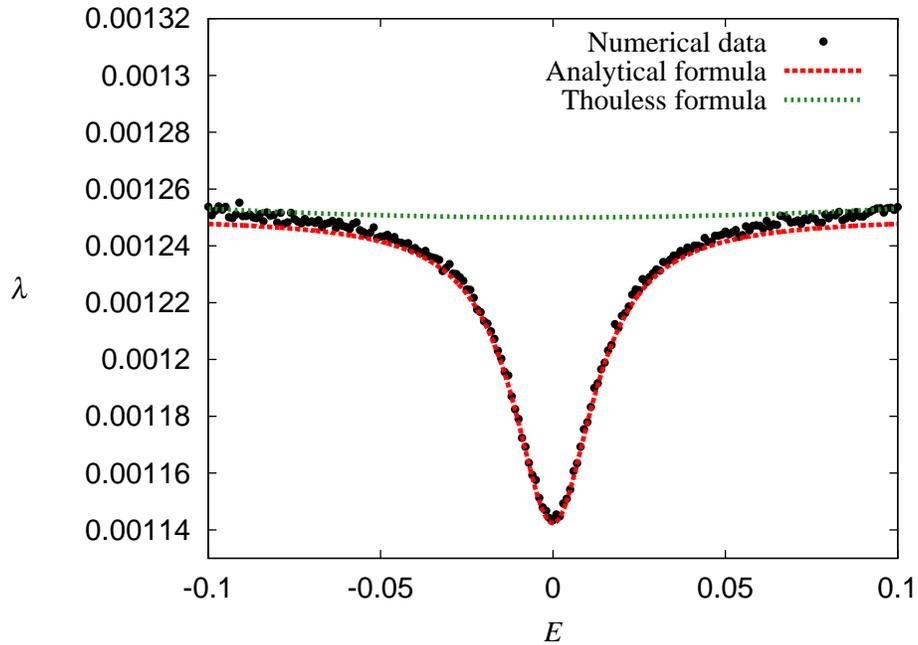}
\caption{\label{lyap} (Colour online)
Inverse localisation length $\lambda$ versus energy $E$. The dots
represent numerically computed values of the Lyapunov exponent; the
dashed line is given by Eq.~(\ref{general}); the dotted line
corresponds to Thouless' formula.} 
\end{center}
\end{figure}

It is not difficult to evaluate numerically Eq.~(\ref{general}); this
operation makes possible to compare analytical and numerical results for
the inverse localisation length.
As shown by Fig.~\ref{lyap}, the theoretical predictions of
Eq.~(\ref{general}) fit very well the numerical data in a neighbourhood
of the band centre.
For the case represented in Fig.~\ref{lyap} the disorder strength was
set at $\sigma^{2} = 0.01$; the considered energy interval $[-0.1,0.1]$
corresponds to values of $\varkappa$ in the range $[-10,10]$.
It is easy to notice that the actual localisation length exhibits
a non-negligible difference from the values predicted by Thouless' formula
even for energies lying quite away from the band centre.
This is a consequence of the slow convergence of the invariant
distribution~(\ref{rholarge}) to the uniform limit discussed in
Sec.~\ref{kappalarge}; we can conclude that the phenomenon of anomalous
localisation is not restricted to an infinitesimal neighbourhood of the
band centre, but can be detected over a finite interval of energy values.

\begin{figure}[t]
\begin{center}
\includegraphics[width=5in,height=3.5in]{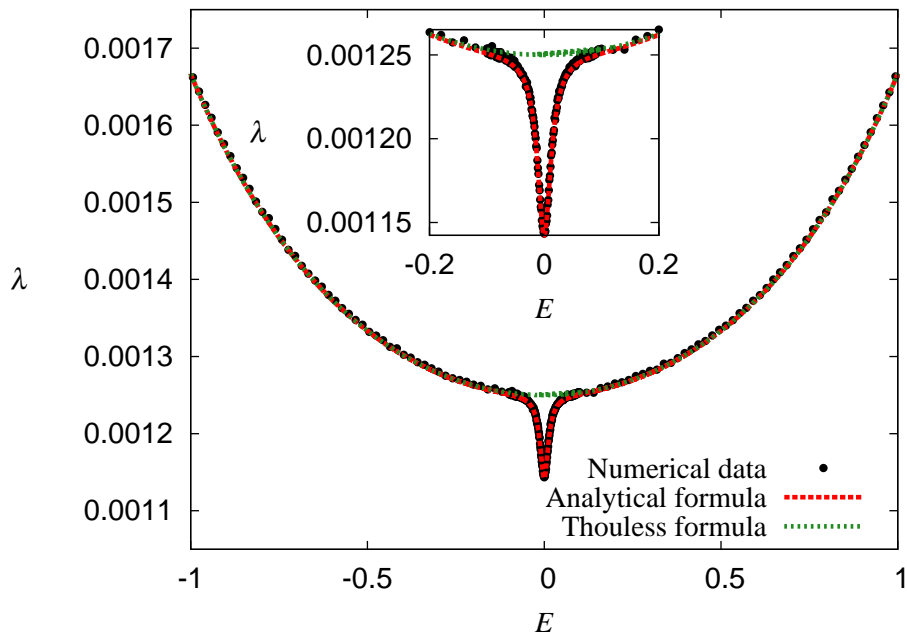}
\caption{\label{interpol} (Colour online)
Inverse localisation length $\lambda$ versus energy $E$. The dots
represent numerically computed values of the Lyapunov exponent; the
dashed line is given by Eq.~(\ref{inter}); the dotted line
corresponds to Thouless' formula. The inset shows in more detail
the same data in the neighbourhood of the band centre.} 
\end{center}
\end{figure}

A close examination of Fig.~\ref{lyap} shows that, for energies
$|E| \gtrsim 0.05$, the numerical data tend to overlap with Thouless'
formula, while Eq.~(\ref{general}) flattens. This is due to the fact
that, as $\varkappa$ increases, Eq.~(\ref{general}) tends to a constant
limit (see Eq.~(\ref{lyaplarge}) below). The physical reason is that
Eq.~(\ref{general}) is valid only as $|E| \to 0$.
It turns out, however, that Eq.~(\ref{inter}) works very well over the
whole energy band. This is born out by Fig.~\ref{interpol}, which
compares the numerical data both with Thouless' formula~(\ref{thouless})
and with Eq.~(\ref{inter}).
That Eq.~(\ref{inter}) works so well over the whole energy band is
not surprising. In fact, the rhs of Eq.~(\ref{inter}) is the product
of Thouless' formula and of a corrective factor. The latter is responsible
for the anomalous behaviour of the localisation length in the neighbourhood
of the band centre, but tends to unity for increasing values of $|E|$.
In this way Thouless' expression is recovered away from the band centre
and the anomaly is taken into account by the corrective factor.

\subsection{Localisation length for $|\varkappa| \ll 1$}

We now turn our attention to the limit cases $|\varkappa| \ll 1$ and
$|\varkappa| \gg 1$.
In the case $|\varkappa| \ll 1$, i.e., when the energy lies very close to
the band centre, one can evaluate the rhs of Eq.~(\ref{lyap1}) with the
approximate invariant distribution~(\ref{rhosmall}). In this way one obtains
\begin{equation}
\lambda = \sigma^{2} \left[\frac{\Gamma \left(\frac{3}{4} \right)}
{\Gamma \left( \frac{1}{4} \right)} \right]^{2}
\left[ 1 + O(\varkappa) \right] ,
\label{lyapsmall}
\end{equation}
in agreement with the known result~\cite{Der84,Gol94,Izr98}.
We remark that Eq.~(\ref{lyapsmall}) can also be written in the
equivalent form
\begin{displaymath}
\lambda = \frac{\sigma^{2}}{4} \left[ 2
\frac{\mathbf{E} \left( \frac{1}{\sqrt{2}} \right)}
{\mathbf{K} \left( \frac{1}{\sqrt{2}} \right)} - 1 \right]
\left[ 1 + O(\varkappa) \right]
\end{displaymath}
which is sometimes used in the literature~\cite{Sar81,Eco06}.
The symbol $\mathbf{E}(k)$ represents the complete elliptic integral
of the second kind, i.e.,
\begin{displaymath}
\mathbf{E}(k) = \int_{0}^{\pi/2} \sqrt{1 - k^{2} \sin^{2} \phi} \;
\mathrm{d} \phi .
\end{displaymath}

\subsection{Localisation length for $|\varkappa| \gg 1$}

In the opposite case, i.e., for $|\varkappa| \gg 1$, the energy, although
close to the band centre in absolute terms, moves away from it on the
energy scale set by the strength of the disorder.
When $|\varkappa| \gg 1$ the invariant distribution takes the
form~(\ref{rholarge}) and the inverse localisation length~(\ref{general})
becomes
\begin{equation}
\lambda = \frac{\sigma^{2}}{8} \left[ 1 - \frac{3}{16 \varkappa^{2}}
+ O \left( \frac{1}{\varkappa^{3}} \right) \right] .
\label{lyaplarge}
\end{equation}
Once more, the result coincides with the formula first derived
by Derrida and Gardner~\cite{Der84,Gol94}.

In the limit $|\varkappa| \to \infty$ the inverse localisation length
tends to the value predicted by Thouless' formula~(\ref{thouless});
as already remarked the transition to the regular limit is not sharp
because of the power-law decay of the anomalous term in
the rhs of Eq.~(\ref{lyaplarge}).
It is important to remark, however, that the limit $|\varkappa| \to \infty$
in Eq.~(\ref{lyaplarge}) can be taken only by letting the intensity of the
disorder $\sigma^{2}$ tend to zero faster than the energy $E$.
Simply increasing $|\varkappa|$ while keeping {\em fixed} the disorder
strength eventually leads to the breakdown of the formula~(\ref{lyaplarge}),
which was derived under the assumption that $E \to 0$.
This can be seen in Fig~\ref{lyap}: for energies $|E| \gtrsim 0.05$
the inverse localisation length tends to move away from the anomalous
expression~(\ref{general}) and closer to Thouless' formula.

\section{Conclusions}
\label{conclu}

In this work we have analysed the anomalous localisation of the
eigenstates of the 1D Anderson model for energies close to the
band centre.
Using the Hamiltonian map approach, we have derived two main
results: the general expression~(\ref{invdis}) of the invariant
distribution for the map~(\ref{map}) and the corresponding
formula~(\ref{inter}) for the inverse localisation length.
The first result shows that the random phase approximation, which
is an essential ingredient of the single-parameter scaling theory,
fails close to the band centre.
The failure is not restricted to a negligible energy range because
the convergence of the distribution~(\ref{invdis}) to the uniform
limit follows a power-law behaviour.

The invariant distribution~(\ref{invdis}) also allowed us to obtain
the analytical expression~(\ref{inter}) of the inverse localisation
length.
We have analytically proved that the general formula~(\ref{inter})
describes the Kappus-Wegner anomaly around the band centre; because
it reduces to Thouless' formula when the energy moves away from the
band centre, Eq.~(\ref{inter}) works well over the whole energy band,
as confirmed by the numerical data.

This paper provides new insight for a critical investigation of the
SPS theory with the Hamiltonian map formalism.
In this work we have focused our attention to the 1D map~(\ref{map})
for the angle variable and in this way we have been able to show that the
random phase approximation, and hence the SPS theory, fail in a finite
energy interval around the band centre.
However, an extension of our analysis to the complete action-angle map
given by Eqs.~(\ref{map}) and~(\ref{jmap}) is required in order to study
the statistical properties of the conductance of a random segment.
In fact, the conductance is related to the action variable, which is
statistically correlated with the angle variable: hence a simultaneous
study of both variables cannot be avoided.
We plan to address this issue in a future work.

\section*{Acknowledgements}

L.T. gratefully acknowledges the support of CONACyT grant No. 150484.
F. M. I. acknowledges support from CONACyT grant No. 161665.


\begin{thebibliography}{10000}

\bibitem{And58} P. W. Anderson, {\it Phys. Rev.} {\bf 109}, 1492 (1958)

\bibitem{Ish73} K. Ishii, {\it Suppl. Prog. Theor. Phys.} {\bf 53}, 77
                (1973)

\bibitem{Izr11} F. M. Izrailev, A. A. Krokhin, N. M. Makarov, {\it 
                arXiv:1110.1762}

\bibitem{Tho79} D. J. Thouless, p.1 in ``La mati\`{e}re mal condens\'{e}e -
                Ill-Condensed Matter'', R.~Balian, R.~Maynard, G.~Toulouse
                eds., North-Holland (Amsterdam) and World Scientific
                (Singapore), 1979

\bibitem{Czy81} G. Czycholl, B. Kramer, A. MacKinnon, {\it Z. Phys. B}
                {\bf 43}, 5 (1981)

\bibitem{Kap81} M. Kappus, F. Wegner, {\it Z. Phys. B} {\bf 45}, 15 (1981)

\bibitem{Sar81} S. Sarker, {\it Phys. Rev. B} {\bf 25}, 4304 (1992)

\bibitem{Der84} B. Derrida, E. Gardner, {\it J. Physique} {\bf 45}, 1283
                (1984)

\bibitem{Bov88} A. Bovier. A. Klein, {\it J. Stat. Phys.} {\bf 51}, 501
                (1988); M. Campanino, A. Klein, {\it Comm. Math. Phys.}
                {\bf 130}, 441 (1990)

\bibitem{Kus93} R. Kuske, Z. Scuss, I. Goldhirsch, S. H. Noskowicz,
                {\it SIAM J. Appl. Math.} {\bf 53}, 1210 (1993)

\bibitem{Gol94} I. Goldhirsch, S. H. Noskowicz, Z. Schuss, {\it Phys. Rev.
                B} {\bf 49}, 14504 (1994)

\bibitem{Sch03} H. Schomerus, M. Titov, {\it Phys. Rev. B} {\bf 67},
                100201(R) (2003)

\bibitem{Hei04} J. Heinrichs, {\it J. Phys. C: Condens. Matter}, {\bf 16},
                7995 (2004)

\bibitem{Dey00}  L. I. Deych, A. A. Lisyansky, B. L. Altshuler,
                {\it Phys. Rev. Lett.} {\bf 84}, 2678 (2000);
                L. I. Deych, M. V. Erementchouk, A. A. Lisyansky,
                B. L. Altshuler, {\it Phys. Rev. Lett.} {\bf 91},
                096601 (2003)

\bibitem{Abr79} E. Abrahams, P. W. Anderson, D. C. Licciardello,
                T.~V.~Ramakrishnan, {\it Phys. Rev. Lett.} {\bf 42}, 673
                (1979)

\bibitem{And80} P. W. Anderson, D. J. Thouless, E. Abrahams, D. S. Fischer,
                {\it Phys. Rev. B}, {\bf 22}, 3519 (1980)

\bibitem{Izr95} F. M. Izrailev, T. Kottos, G. Tsironis, {\it Phys. Rev. B}
                {\bf 52}, 3274 (1995)

\bibitem{Izr98} F. M. Izrailev, S. Ruffo, L. Tessieri, {\it J. Phys. A:
                Math. Gen.} {\bf 31}, 5263 (1998)

\bibitem{Tes00} L. Tessieri, F. M. Izrailev, {\it Phys. Rev. E} {\bf 62},
                3090 (2000); L. Tessieri, F. M. Izrailev, {\it Phys. Rev. E}
                {\bf 64}, 066120 (2001)

\bibitem{Lam82} C. J. Lambert, M. F. Thorpe, {\it Phys. Rev. B} {\bf 26},
                4742 (1982); A. D. Stone, D. C. Allan, J. D. Joannopoulos,
                {\it Phys. Rev. B} {\bf 27}, 836 (1983)

\bibitem{Kan11} Kai Kang, Shaojing Qin, Chuilin Wang, {\it Phys. Lett. A}
                {\bf 375}, 3529 (2011)

\bibitem{Har60} G. H. Hardy, E. M. Wright, {\it An Introduction to the
                Theory of Numbers}, 4th ed., Oxford University Press,
                Oxford (1960)

\bibitem{Gar04} C. W. Gardiner, {\it Handbook of Stochastic Methods},
                3rd ed., Springer Verlag, Berlin (2004)

\bibitem{Eco06} E. N. Economou, {\it Green's Functions in Quantum
                Physics}, 3rd ed., Springer, Berlin (2006)


\end{thebibliography}
\end{document}